\documentstyle[11pt,newpasp,twoside,psfig,epsf]{book}

\newcommand{\ie}{{\em i.e.},\ }
\newcommand{\eg}{{\em e.g.},\ }
\newcommand{\etal}{{\em et al.}\ }
\newcommand{\cf}{{\em cf.}\ }
\newcommand{\etc}{{\em etc.}\ }

\begin{document}





\pagenumbering{arabic}
\setcounter{page}{52}

\markboth{Djorgovski, \textit{et al}.}{Rare Objects}

\setcounter{section}{0}
\setcounter{figure}{0}
\setcounter{table}{0}
\setcounter{footnote}{0}

\title{Searches for Rare and New Types of Objects}

\author{S.G. Djorgovski, A.A. Mahabal, R.J. Brunner, R.R. Gal, S. Castro}
\affil{Palomar Observatory, California Institute of Technology, Pasadena, CA, 91125}
\author{R.R. de Carvalho}
\affil{Observatorio Nacional, CNPq, Rio de Janeiro, Brasil}
\author{S.C. Odewahn}
\affil{Department of Physics \& Astronomy, Arizona State University, Tempe, AZ, 85287}

\begin{abstract}
Systematic exploration of the observable parameter space, covered by
large digital sky surveys spanning a range of wavelengths, will be one
of the primary modes of research with a Virtual Observatory (VO).
This will include searches for rare, unusual, or even previously
unknown types of astronomical objects and phenomena, \eg as outliers
in some parameter space of measured properties, both in the catalog
and image domains.  Examples from current surveys include
high-redshift quasars, type-2 quasars, brown dwarfs, and a small
number of objects with puzzling spectra.  Opening of the time domain
will be especially interesting in this regard.  Data-mining tools such
as unsupervised clustering techniques will be essential in this task,
and should become an important part of the VO toolkit.
\end{abstract}

\section{Introduction: Mining the Sky}

The great quantitative increases in the amount and complexity of
information harvested from large digital sky surveys, with information
volumes now measured in multiple Terabytes (and soon Petabytes), with
billions of sources detected and tens or hundreds of parameters
measured for each of them, pose some fundamental questions: Will this
quantitative increase lead to a qualitative change in the way we do
astronomy?  Will we start asking new kinds of questions about the
universe, and use new methods in answering them?  How to exploit this
great riches of information in a systematic and effective way, and how
to extract the scientific essence and knowledge from this mass of bits
and pixels?  This is indeed what a Virtual Observatory (VO) idea is
all about.

There may be two (or, better yet, at least two) main streams of the
new, VO-enabled astronomy:

First, there will be statistical astronomy ``done right'', \ie studies
such as the mapping and quantification of the large scale structure in
the universe, of the Galactic structure, construction and studies of
complete samples of all kinds of objects (stars or galaxies of
particular types or particular ranges of properties, AGN, clusters of
galaxies, \etc).  This is the ``bread-and-butter'' of astronomy, the
way to map and quantify our universe in a systematic, statistically
sound fashion, and to feed and constrain our basic theoretical models
and understanding.  We should never again be limited by the Poissonian
errors from small samples of objects; of course, understanding of
possible systematic errors and biases in the sky surveys now becomes
even more important. Both the numbers of sources and the wide-angle
coverage are important for such studies.  In some sense, this will be
a direct extrapolation of the type of astronomy we have been doing all
along, but brought to a higher level of accuracy and detail by the
sheer information content of the new, digital sky.

The second stream, where we may expect more novelty and surprises, is
a systematic exploration of the poorly known portions of the
observable parameter space, and specifically searches for rare types
of astronomical objects and phenomena, both already known, and as yet
unknown.  Here we can use the large numbers of detected sources to
look for rare events which would be unlikely to be found in smaller
data sets: if some type of an interesting object is, say one in a
million or a billion down to some flux limit, then we need a sample of
sources numbering in many millions or billions in order to discover a
reasonable sample of such rare species.  Rare objects may be
indistinguishable from the more common varieties in some observable
parameters (\eg quasars look just like normal stars in images), but be
separable in other observable axes (\eg the shape of the broad-band
spectral energy distribution).  This type of new astronomy with large
digital sky surveys (and a VO) is the subject of this review.

\section{Exploring the Parameter Space}

Some axes of the observable parameter space are obvious and well
understood: the flux limit (depth), the solid angle coverage, and the
range of wavelengths covered.  Others include the limiting surface
brightness (over a range of angular scales), angular resolution,
wavelength resolution, polarization, and especially variability over a
range of time scales; all of them at any wavelength, and again as a
function of the limiting flux.  In some cases (\eg the Solar system,
Galactic structure) apparent and proper motions of objects are
detectable, adding additional information axes.  For well-resolved
objects (\eg galaxies), there should be some way to quantify the image
morphology as one or more parameters.  And then, then there are the
non-electromagnetic information channels, \eg neutrinos, gravity
waves, cosmic rays \ldots The observable parameter space is enormous.

We can thus, in principle, measure a huge amount of information
arriving from the universe, and so far we have sampled well only a
relatively limited set of sub-volumes of this large parameter space,
much better along some axes than others: We have fairly good sky
surveys in the visible, NIR, and radio; more limited all-sky surveys
in the x-ray and FIR regimes; \etc For example, it would be great to
have an all-sky survey at the FIR and sub-mm wavelengths, reaching to
the flux levels we are accustomed to in the visible or radio surveys,
and with an arcsecond-level angular resolution; this is currently
technically difficult and expensive, but it is possible.  The whole
time domain is another great potential growth area.  Some limits are
simply technological or practical (\eg the cost issues); but some are
physical, \eg the quantum noise limits, or the opacity of the Galactic
ISM.

Historically, the concept of the systematic exploration of the
universe through a systematic study of the observable parameter space
was pioneered by Fritz Zwicky, starting in 1930's (see, \eg Zwicky
1957).  While his methodology and approach did not find many
followers, the core of the important ideas was clearly there.  Zwicky
was limited by the technology available to him at the time; probably
he would have been a major developer and user of a VO today!  Another
interesting approach was taken by Harwit (1975; see also Harwit \&
Hildebrand 1986), who examined the limits and selection effects
operating on a number of axes of the observable parameter space, and
tried to estimate the number of fundamental new (``class A'')
astrophysical phenomena remaining to be discovered.  While one could
argue with the statistics, philosophy, or details of this analysis, it
poses some interesting questions and offers a very general view of our
quest to understand the physical universe.

So, it is not just the space we want to study; it is the parameter
space (in the cyber-space).  Much of the total observable parameter
space which is in principle (\ie technologically) available to us is
still very poorly sampled.  This is our {\sl Terra Incognita}, which
we should explore systematically, and where we have our best chance to
uncover some previously unknown types of objects or astrophysical
phenomena --- as well as reach a better understanding of the already
known ones.

This is an ambitious, long-term program, but even with a relatively
limited coverage of the observable parameter space we already have in
hand it is possible to make some significant advances.

\section{Looking for the Rare, but Known Types of Objects}

Some types of astronomical objects, \eg particular types of stars or
quasars may be relatively rare, or simply be hard to find in the
available data sets.  But we could use some of their known or expected
properties (\eg typical broad-band spectra, or variability) folded
through the survey selection functions (\eg bandpass curves) to design
experiments where such objects can be distinguished from the
``uninteresting'' majority (\eg normal stars or galaxies).  This
approach has been used very successfully in the past: most quasars
have been found using some such approach, first as ``radio-loud
stars'', then as UV excess objects, \etc; ultraluminous IRAS galaxies
have anomalously large FIR/visible flux ratios; variable stars and
distant supernov\ae\ distinguish themselves with particular types of
light curves; and so on.

Sometimes a simple cross-wavelength match can reveal interesting
objects or phenomena by indicating those with unusual broad-band
energy distributions: recall the discovery of quasars and
radio-galaxies, or ULIRGs, or LMXBs, or intra-cluster x-ray gas, or
the recent progress on GRBs through the study of their afterglows.
This is an obvious area where a VO can be used to construct a
detailed, panchromatic view of the universe, and isolate different
kinds of objects, with better understanding of the observational
biases and selection effects; Lonsdale's contribution to this volume
illustrates such an approach to a general census of AGN.

If objects are spatially unresolved in some sky survey, then the only
distinguishing information is in their flux ratios between different
bands, \eg colors.  As an example, FIR flux ratios have been used to
classify IRAS sources as probable stars or galaxies (\eg Boller \etal
1992).

Even within a given survey with a limited wavelength baseline this
approach can be used to separate physically distinct types of objects,
or, through some photometric redshift indicator, objects of a given
type but in different redshift ranges.  This color selection technique
is now the principal discovery method for quasars at $z \ga 4$
(Warren \etal 1987, Irwin \etal 1991, Kennefick \etal 1995a, 1995b,
Fan \etal 1999, 2000a, 2000c, \etc), or brown (L and T) dwarfs
(Kirkpatrick \etal 1999, Strauss \etal 1999, Burgasser \etal 2000, Fan
\etal 2000b, Leggett \etal 2000, \etc).

As an illustration, in Figure 1 we show how the color selection works
with the examples of high-$z$ and type-2 quasars discovered in DPOSS
(Djorgovski \etal 1998; and in preparation).  Normal stars form a
temperature sequence, seen here as a banana-shaped locus of points in
the parameter space of colors.  The spectra of these quasars, when
folded through the survey filter curves (Figures 2 and 3), produce colors
discrepant from those of normal stars.

\begin{figure}[!htb]
\plotfiddle{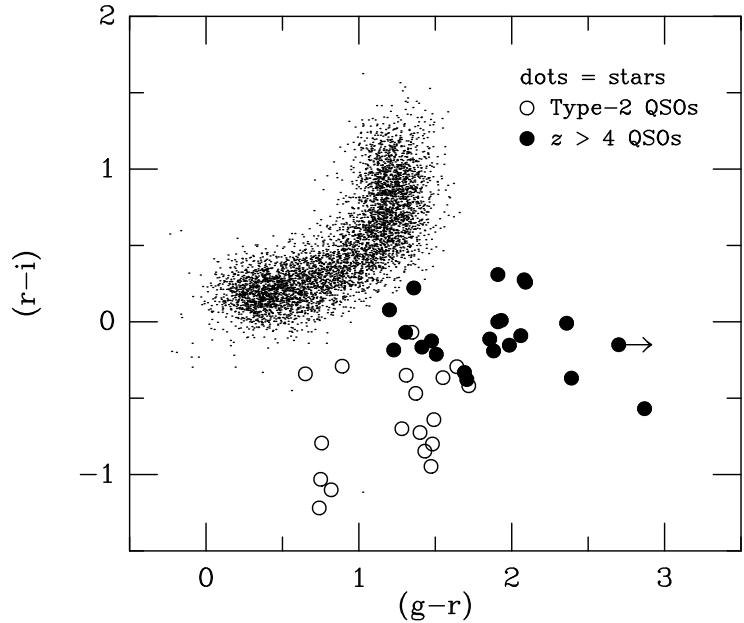}{3.5in}{0}{75}{75}{-180}{-36}
\caption{
A representative color-color plot for objects classified as PSF-like
in DPOSS.  The dots are normal stars with $r \sim 19$ mag.  Solid
circles are some of the $z > 4$ quasars, and open circles are some of
the type-2 quasars found in this survey.  While quasars are
morphologically indistinguishable from ordinary stars, this color
parameter space offers a good discrimination among these types of
objects.  Similar methodology is now also used to discover brown
dwarfs in SDSS and 2MASS.}
\end{figure}

In the case of high-$z$ quasars, absorption by the intergalactic
hydrogen (the Ly$\alpha$ forest) produces a strong drop blueward of
the quasar's own Ly$\alpha$ emission line center, and thus a very red
$(g-r)$ color, while the observed $(r-i)$ color reflects the
intrinsically blue spectrum of the quasars: these objects are red in
the blue part of the spectrum, and blue in the red part of the
spectrum --- unlike any stars.  To date, $\sim 100$ such quasars have
been found in DPOSS; we make them publicly available through our
webpage\footnote{http://www.astro.caltech.edu/$\sim$george/z4.qsos}.
At intermediate Galactic latitudes, there is about one of them per
million stars, down to $r \sim 19.5$ magnitude.  Thus a good color
discrimination and a good star-galaxy separation are essential in
order to avoid an excessive contamination of the spectroscopic
follow-up samples by mismeasured stars or misclassified galaxies.  A
variant of this technique (based also on the Lyman-limit drop) is now
used to find galaxies at $z \ga 3$ (\eg Steidel \etal 1999, Dickinson
\etal 2000, and references therein).

\begin{figure}[!htb]
\plotfiddle{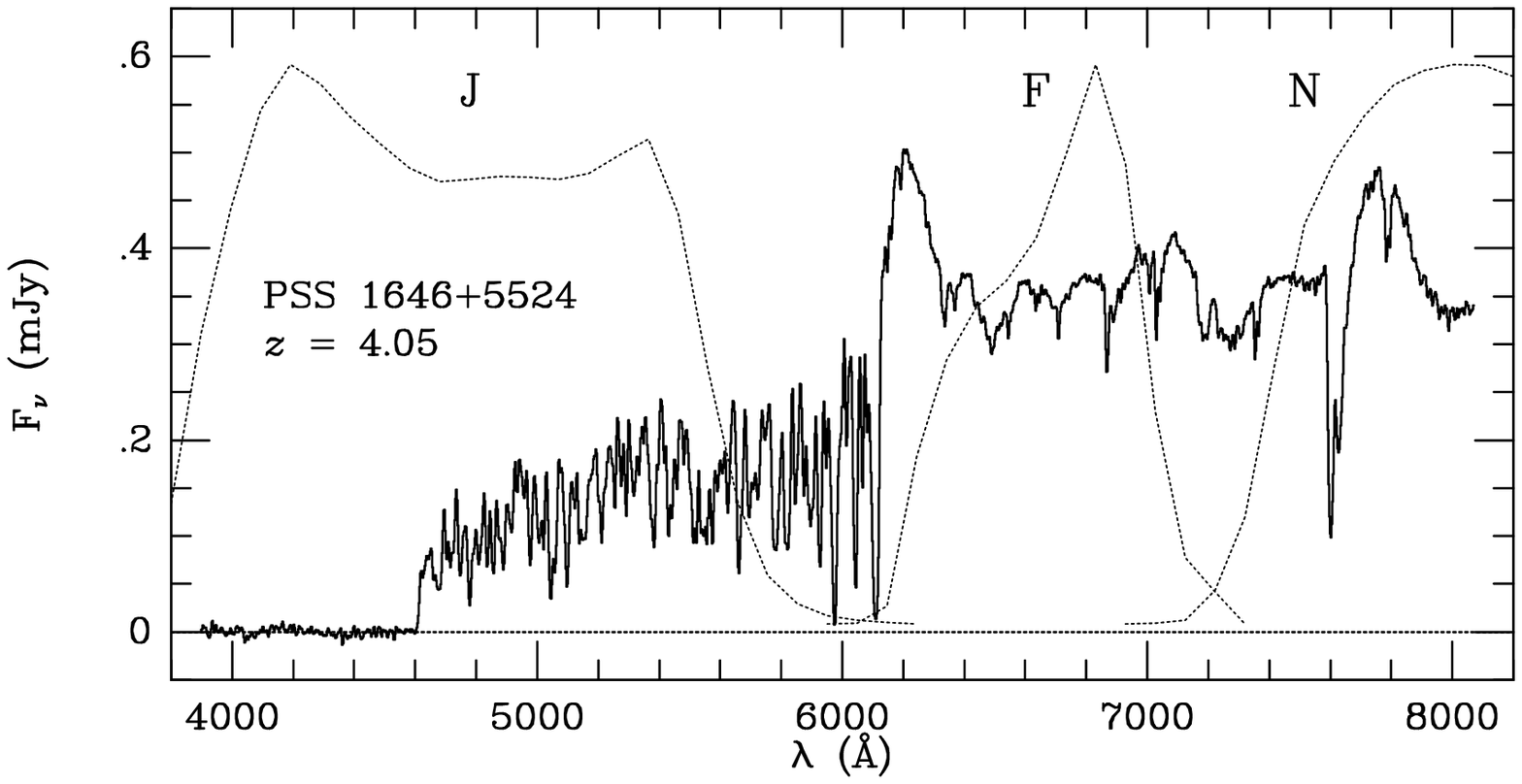}{2.5in}{0}{75}{75}{-216}{-45}
\caption{
A spectrum of a typical $z > 4$ quasar, with the DPOSS bandpasses shown as
dotted lines.  The mean flux drop blueward of the Ly$\alpha$ line, caused by
the absorption by Ly$\alpha$ forest and sometimes a Lyman-limit system, gives
these objects a very red $(g-r)$ color, while their intrinsic blue color is
retained in $(r-i)$.  This places them in the portion of the color parameter
space indicated in Figure 1. 
}
\end{figure}

A similar ``convex spectrum'' effect can be caused by the presence of
strong emission lines in the middle band ($r$), as shown here in the
example of type-2 quasars discovered in DPOSS (Djorgovski \etal 1999,
and in prep.).  We found a whole population of these long-sought
objects (which are now also appearing in considerable numbers in the
CXO x-ray data), selected through their peculiar colors.  The
selection effects are complex, depending both on the [O III] line
fluxes and equivalent widths, so we find only a subset of them, those
with a mostly unobscured narrow-line region, in the redshift interval
given by the width of the DPOSS $r$ band ($z \sim 0.31$--$0.38$ for the
[O III] lines).  These objects are sufficiently rare, with surface
density $\la 10^{-2}$ per square degree for our selection criteria, that
one must have a survey covering a very large area, yet go sufficiently
deep to detect the host galaxies (in our survey most of the light in
the $g$ and $i$ bands is from the hosts).  This is why this population
was missed in the past, with surveys lacking either the necessary
depth or the area coverage.

\begin{figure}[!htb]
\plotfiddle{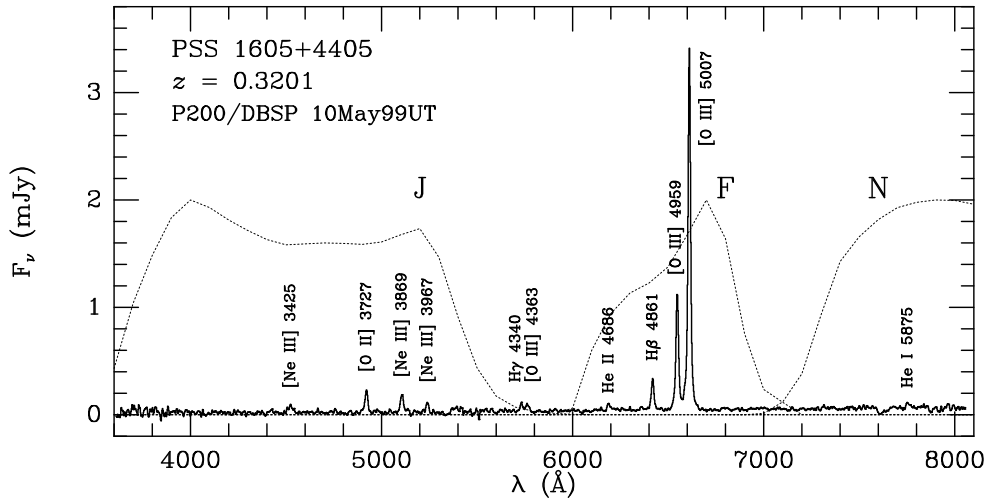}{2.5in}{0}{75}{75}{-216}{-45}
\caption{
A spectrum of a typical type-2 quasar, with the DPOSS bandpasses shown
as dotted lines.  The presence of the strong [O III] lines is the $r$
band places such objects in the portion of the color parameter space
indicated in Figure 1.  }
\end{figure}

This simple, but very efficient and demonstrably successful method can
be used to isolate other kinds of sources as well, \eg stars of a
particular spectral type, to be used as tracers of Galactic structure,
or the samples of mostly unobscured quasars in general (\cf Wolf \etal
1999 or Warren \etal 2000).  Multiplicity of bandpasses and the
dynamical range of the wavelength baseline help; after all, multicolor
photometry can be viewed as an extremely low resolution spectroscopy.

Analogous techniques could be used in other parameter spaces, for example for
an objective classification and selection of galaxies of a particular type,
when image morphology can be quantified appropriately.

\section{Looking for New Kinds of Objects}

Perhaps the most intriguing new scientific prospect for a VO is the
possibility of discovery of previously unknown types of astronomical
objects and phenomena.  Such things might have been missed so far
either because they are rare, or because they would require a novel
combination or a way of looking at the data.  A thorough, large-scale,
unbiased, multi-wavelength census of the universe will uncover them,
if they do exist (and surely we have not yet found all there is out
there).  Methodology similar to that used to find known rare types of
objects, \ie as outliers in some suitably chosen, discriminative
parameter space, can be used to search for the possible new species.
This ``organized serendipity'' can lead to some exciting new
discoveries.

Possible examples of new kinds of objects (or at least extremely rare
or peculiar sub-species of known types of objects) have been found in
the course of high-$z$ quasar searches by both SDSS (Fan \& Strauss,
private communication) and DPOSS groups.  Two examples from DPOSS are shown in
Figures 4 and 5.  These objects have most unusual, and as yet not fully (or
not at all) understood spectra, which cause them to have peculiar
broad-band colors. Their colors places them in the designated portions
of the color space where high-$z$ quasars are to be found, and clearly
other, as yet unexplored portions of this parameter space may contain
additional peculiar objects.  While some may simply turn out to be
little more than curiosities, others may be representative of genuine
new astrophysical phenomena.

\begin{figure}[!htb]
\plotfiddle{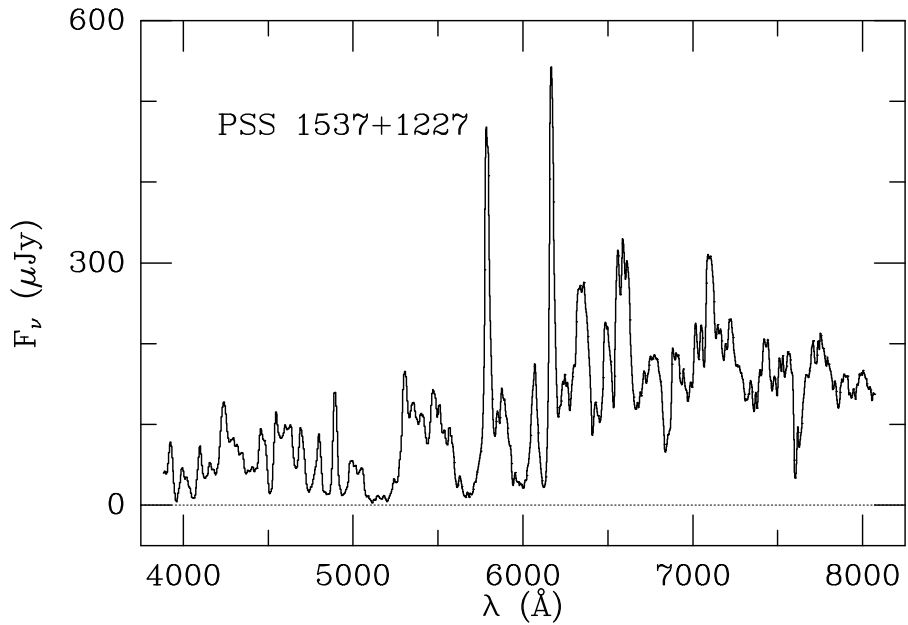}{3.5in}{0}{75}{75}{-216}{-45}
\caption{
A spectrum of a peculiar object PSS 1537+1227, obtained at Palomar.  The object
was initially selected as a high-$z$ quasar candidate due to its colors, in a
manner illustrated in Figure 1.  It turned out to be an extreme case of a rare
type of a low-ionization, Fe-rich, BAL QSO, at $z \approx 1.2$.  A prototype
case (but with a spectrum not quite as extreme as this) is FIRST 0840+3633, 
discovered by Becker \etal (1997).
}
\end{figure}

\begin{figure}[!htb]
\plotfiddle{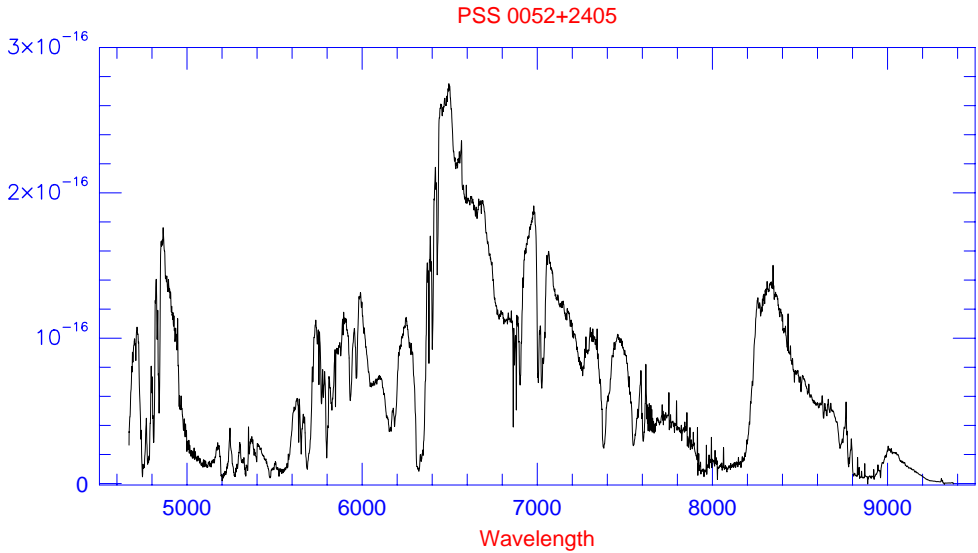}{3.0in}{0}{55}{55}{-200}{-81}
\caption{
A spectrum of another peculiar object, PSS 0052+2405, obtained at Keck.  This
object was also initially selected as a high-$z$ quasar candidate.  Its nature
is still uncertain, but it may be another example of a peculiar BAL QSO --- or
something completely different.
}
\end{figure}

In order to tackle this problem right, we need proper computational
and statistical tools, generally falling in the area of unsupervised
clustering or classification, which is a part of the more general and
rapidly growing field of Data Mining (DM) and Knowledge Discovery in
Databases (KDD).  This opens up great opportunities for collaborations
with computer scientists and statisticians.  For an overview of some
of the issues and methods, see the volume edited by Fayyad \etal
(1996b), as well as several papers in this volume.  Good visualization
tools are also essential for this task.

If applied in the catalog domain, the data can be viewed as a set of
$n$ points or vectors in an $m$-dimensional parameter space, where $n$
can be in the range of many millions or even billions, and $m$ in the
range of a few tens to hundreds.  The data may be clustered in $k$
statistically distinct classes, which could be modeled, \eg as
multivariate Gaussian clouds in the parameter space, and which
hopefully correspond to physically distinct classes of objects (\eg
stars, galaxies, quasars, \etc).  This is a computationally highly
non-trivial problem, approaching Terascale supercomputing, and it
calls for some novel and efficient implementations of clustering
algorithms.  However, not all parameters may be equally interesting or
discriminating, and lowering this dimensionality to some more
appropriate subset of parameters would be an important task for the
scientists actually using such tools to explore the data.

If the number of object classes $k$ is known (or declared) {\it a
priori}, and training data set of representative objects is available,
the problem reduces to supervised classification, where tools such as
Artificial Neural Nets (ANN) or Decision Trees (DT) can be used.  This is now
commonly done for star-galaxy separation in the optical or NIR sky
surveys (\eg Odewahn \etal 1992, or Weir \etal 1995), and searches for
known types of objects with predictable signatures in the parameter
space (\eg high-$z$ quasars) can be also cast in this way.

However, a more interesting and less biased approach is where the
number of classes $k$ is not known, and it has to be derived from the
data themselves.  The problem of unsupervised classification is to
determine this number in some objective and statistically sound
manner, and then to associate class membership probabilities for all
objects.  Majority of objects may fall into a small number of classes,
\eg normal stars or galaxies. What is of special interest are objects
which belong to much less populated clusters, or even individual
outliers with low membership probabilities for any major class.  Some
initial experiments with unsupervised clustering algorithms in the
astronomical context include, \eg Goebel \etal (1989), Weir \etal
(1995), de Carvalho \etal (1995), and Yoo \etal (1996), but a
full-scale application to major digital sky surveys yet remains to be
done.  An array of good unsupervised classification techniques will be
an essential part of a VO toolkit.

\section{Other Domains of the Parameter Space}

Most of the work described so far involved searches in the catalog
domain, and specifically in the parameter spaces of colors measured in
optical and NIR sky surveys.  However, many other domains of the
observable parameter space are still wide open and waiting to be fully
explored.

The low surface brightness universe (at any wavelength!) is one of the
obvious frontiers, and is addressed elsewhere in this volume by
Schombert and by Brunner \etal; see also the review by Impey \& Bothun
(1997), and references therein.  Conversely, we may be missing some
compact, {\em high} surface brightness galaxies (a possibility
envisioned by Zwicky many decades ago): \cf Drinkwater \etal (1999);
however, a field spectroscopic survey of almost-unresolved DPOSS
objects at Palomar (Odewahn \etal, in prep.) failed to turn up a
substantial number of such objects.  In any case, expanding the
dynamical range of the limiting surface brightness and angular
resolution in digital sky surveys at any wavelength is likely to be
one of the key area of research in a VO.

Perhaps the most promising new domain for exploration is the time
domain: variability at all time scales, and all wavelengths, be it
periodic, eruptive, or chaotic in nature.  The subject is addressed by
Diercks elsewhere in this volume, and by Paczy\'nski (2000).  The
importance and the scientific promise of the exploration of the time
domain has been recognized through the high recommendation of the NAS
Decadal Report, {\sl Astronomy and Astrophysics in the New
Millennium}, of the Large Synoptic Survey Telescope (LSST).  Other
large-scale sky monitoring program are already in progress (\eg
Akerlof \etal 2000, Groot \etal 2000, Everett \etal 2000, and the many
searches for the Solar system objects reviewed by Pravdo elsewhere in
this volume). Synoptic monitoring of the sky over a range of
wavelengths, and mining of the resulting multi-Petabyte data sets may
be the most technically demanding and among the most scientifically
productive areas for a VO.

\begin{figure}[!htb]
\plotfiddle{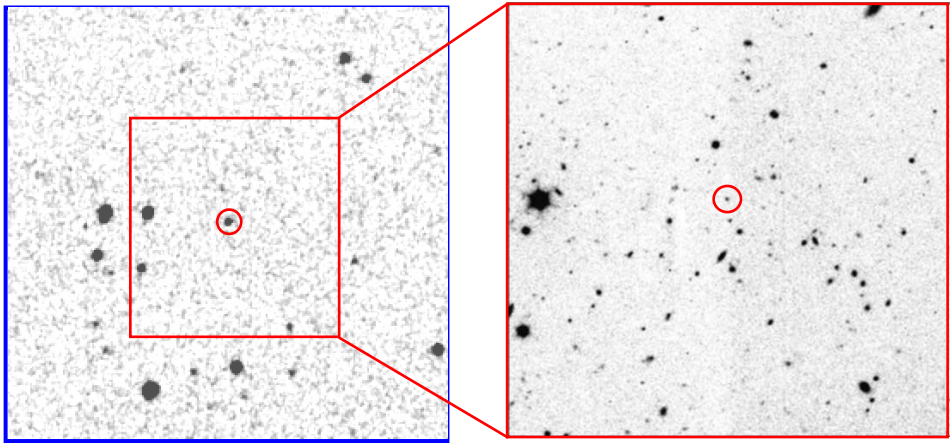}{3.0in}{90}{50}{50}{189}{-63}
\caption{
An example of a serendipitously discovered optical transient event
from DPOSS.  The left panel shows a portion of a DPOSS $F$ plate image
with an $r \sim 18.5$ magnitude, starlike object circled.  The object was
selected due to its apparent peculiar color (bright in $r$, extremely
faint in the other two DPOSS bands); however, this was simply a
consequence of the plates taken at different times, with one of them
catching it in a bright state.  The right panel shows a portion of
the corresponding Keck $R$ band image.  The DPOSS transient was
positionally coincident with an $R \sim 24.5$ magnitude galaxy, with an
estimated probable $z \sim 1$.  At such a redshift, this object would
have been a few hundred times brighter than a supernova at its peak.
It may be an example of a GRB ``orphan afterglow'', or possibly some
other, new type of a transient.  }
\end{figure}

Most of the studies described so far involve searches in some
parameter or feature space, \ie catalogs derived from survey images.
However, we can also contemplate a direct exploration of sky surveys
in the image (pixel) domain.  Automated pattern recognition and
classification tools can be used to discover sources with a particular
image morphology (\eg galaxies of a certain type). An example from
planetary science, an automated discovery of volcanos in Magellan
Venus radar images, was described in Fayyad \etal (1996a) and Burl
\etal (1998).  An even more interesting approach would be to employ AI
techniques to search through panoramic images (perhaps matched from
multiple wavelengths) for unusual image patterns.  For example, it may
be possible for a program to discover gravitationally lensed arcs in rich
clusters, and possibly some other, as yet unknown phenomena.

Finally, an unsupervised classification search for unusual patterns or
signals in astronomical data represents a natural generalization of
the SETI problem (Djorgovski 2000).

\section{Concluding Comments}

A VO, applied on the plethora of large, digital sky surveys would
enable a thorough and systematic exploration of the observable
parameter space, leading to a more complete understanding of the
physical universe.  Introduction of novel DM and KDD techniques,
developed in collaboration with computer scientists will be essential.
In addition to the construction of significant samples of various rare
types of astronomical objects which can be used for further studies,
we are likely to find some completely new things.  In this way, a VO
will be a $unique$ tool of astronomical discovery.

There is, however, one significant bottleneck which we can already
anticipate in this type of studies: the follow-up spectroscopy of
interesting sources selected from imaging surveys.  While there seems
to be a vigorous ongoing and planned activity to map and monitor the
sky in many ways and many wavelengths, spectroscopic surveys will be
necessary in order to interpret and understand the likely
overabundance of interesting objects found.  This is something we have
to consider in our plans.

\acknowledgements

The processing and initial exploration of DPOSS was supported by a
generous gift from the Norris foundation, and by other private donors.
Prototyping VO developments at Caltech and JPL have been funded by
grants from NASA, the Caltech President's Fund, and several private
donors.  We are grateful to all people who helped with the creation of
DPOSS and our Palomar and Keck observing runs, and especially a number
of excellent Caltech undergraduates who worked with us through the
years.  Work on the applications of machine learning and AI technology
for exploration of large digital sky surveys was done in collaboration
with U. Fayyad, P.  Stolorz, R. Granat, A. Gray, J. Roden,
D. Curkendall, J. Jacob, and others at JPL.

\end{document}